# GENETIC INTERACTIONS FROM FIRST PRINCIPLES

J Fernandez-de-Cossio[1]*, J Fernandez-de-Cossio-Diaz[2], Y Perera[3].
[1]Bioinformatics Department, Center for Genetic Engineering and Biotechnology, PO Box 6162, CP10600, Havana (Cuba).
[3]Molecular Oncology Group, Pharmaceutical Division, CIGB.
[2]Systems Biology Department, Center of Molecular Immunology, PO Box 6162, CP10600, Havana (Cuba).
*jorge.cossio@cigb.edu.cu

## SHORT TITLE
Revisiting genetic interaction.

## ABSTRACT

We derive a general statistical model of interactions, starting from probabilistic principles and elementary requirements. Prevailing interaction models in biomedical researches diverge both mathematically and practically. In particular, genetic interaction inquiries are formulated without an obvious mathematical unity. Our model reveals theoretical properties unnoticed so far, particularly valuable for genetic interaction mapping, where mechanistic details are mostly unknown, distribution of gene variants differ between populations, and genetic susceptibilities are spuriously propagated by linkage disequilibrium. When applied to data of the largest interaction mapping experiment on Saccharomyces Cerevisiae to date, our results imply less aversion to positive interactions, detection of well-documented hubs and partial remapping of functional regions of the currently known genetic interaction landscape. Assessment of divergent annotations across functional categories further suggests that positive interactions have a more important role on ribosome biogenesis than previously realized. The unity of arguments elaborated here enables the analysis of dissimilar interaction models and experimental data with a common framework.

## INTRODUCTION
The underlying mechanisms of complex functional networks are typically poorly understood or inaccessible to measurements (*1*, *2*). The available information for wiring these interaction networks typically consists of incidence patterns of factors at one end of the causation process, with the outcome of an inquired effect at the other far end. These patterns can indicate the presence of functional interconnected events, though they rarely provide cues of when, where and how the intermediate events interconnect. However, when the immediate consequences of factors are separated in space-time, they are often regarded as non-interacting, even if their farther consequences interconnect before reaching the effect. Evading the notion of "distant" interactions leads to the recommendation of multiple models for no-interaction (null-models) within the same study (*3–9*). For example, besides the risk-additive null-model vehemently advocated in epidemiology, a risk-multiplicative null-model has been invoked for factors acting separately on different steps of a multistage pathogenesis process(*3*, *5*), and has been also prescribed for independence of protecting factors(*4*, *6–9*).

A multiplicity of no-interaction models conveys unnecessary conceptual and pragmatic difficulties. How many non-interaction null models can be conceived within the same study? Which model is appropriate in each tested pair? Can an interaction model prescribed for one case be mathematically equivalent to a non-interaction null-model prescribed for another case? How to distinguish one from the other in each particular case? Space-time resolution of connected events are hardly possible from incidence data of factors and effect only.

A more practical perspective has been tacitly conveyed in large scale interaction studies by relaying on a single non-interaction model for the whole analysis(*10–12*). We learn as much as the evidence permits, by the sole diagnostic of the presence of interactions, disregarding their space-time separation. However, further conceptual ambivalences hinder the identification of a unique appropriate model. It is ussually agreed that factors interact when the effect of their combination deviate from what would be expected based on their individual separate effects (*4*, *13*). Though intuitive, far from guiding to definite physical meanings this consensus merely defers the issue to what a reasonable expectation of the effect of joint independent factors is(*2*, *14*, *15*), leaving the resolution of the concept open. Indeed, diverse and inconsistent measures of interaction pervade the literature (*7*, *10–13*, *16–20*).

Here we undertake a novel resolution of the concept of interaction from basic probabilistic principles and elementary requirements, with a definition that embraces distant interactions. The stated requirements are enough to determine a unique model. Our derivation reveals theoretical properties unnoticed so far, which turned into superior performance when analyzing the data of the largest wiring experiment on *Saccharomyces Cerevisiae* to date(*12*).

## RESOLUTION OF THE CONCEPT OF INTERACTION
### Probabilistic Model

We are interested in classifying the interaction or non-interaction relationships between factors with respect to an effect. The contextual background is generally unknown, and the effects are not fully under control of the factors status, responding stochastically. A probabilistic framework is required to account for this uncertainty.

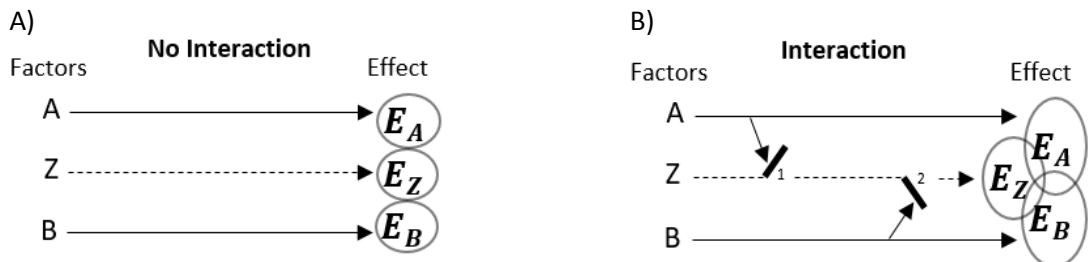

*Figure 1: Sketch of interaction and no-interaction scenarios. A and B are observed factors, Z accounts for unobserved process, and E is the effect of interest. The effect is classified into variants $E_A$, $E_B$ and $E_Z$, related to each factor. a) No-interaction scenario. b) Interaction scenario (there is a cross-over of the pathways from factors leading to the effect).*

Figure 1 sketches response patterns of mechanisms perturbed by two binary observed factors $A$ and $B$ leading to a dichotomous effect $E$. Unknown background agents are considered in the unobservable $Z$. The occurrence of $E$ can be classified into three manifestations $E_A$, $E_B$ and $E_Z$, corresponding to the effect of exposure to factors $A$, $B$ and $Z$ respectively (Figure 1). This correspondence can be represented by the logical expression $E \equiv E_A \vee E_B \vee E_Z$, where ∨ denotes logical OR. Note that $E_A$, $E_B$ and $E_Z$ are each other not necessarily exclusive. Generally, our data does not allow us to discern between these manifestations, undistinguishably observable as $E$. This decomposition is only a temporary construct. As we will see, the variants $E_A$, $E_B$ and $E_Z$ cancel in our derivations, leaving equations that depend only on observables.

Our elementary assumption is that for non-interacting factors $A$ and $B$ the probability $\Pr(\overline{E}|x\ y) = 1 - \Pr(E|x\ y)$ must satisfy the following factorization:

$$\Pr(\overline{E}|x\,y) = \Pr(\overline{E_A}|x)\Pr(\overline{E_B}|y)\Pr(\overline{E_Z}), \qquad \forall x \in A, y \in B \tag{1}$$

for all realizations $x$, $y$ of the factors $A$, $B$. The overbar denotes logical negation, henceforth indicating the non-occurrence of an effect or the non-exposure to a factor. Concatenation denotes logical AND. In this case $\overline{E} = \overline{E_A}\,\overline{E_B}\,\overline{E_Z}$, since none of $E_A$, $E_B$ nor $E_Z$ is realized when the effect $E$ is not realized. A particular logical framework yielding ( 1 ) is shown in Methods. Everything that follows is derived from ( 1 ), which is the minimal requirement of no-interaction.

## Neutral model

Factorization ( 1 ) has practical limitations since it is expressed in terms of the un-observables $E_A$, $E_B$ and $E_Z$. It is shown in Methods that the probability $\Pr(\overline{E}|x\,y)$ of non-interacting factors $A$ and $B$ must satisfy

$$\Pr(\overline{E}|x\,y) = \mathcal{N}(x,y), \qquad \forall x \in A, y \in B \tag{2}$$

where the function $\mathcal{N}(x,y)$ is defined as

$$\mathcal{N}(x,y) = \frac{\Pr(\overline{E}|x)\Pr(\overline{E}|y)}{\sum_{a \in A, b \in B} \Pr(\overline{E}|a\,b)\Pr(a|y)\Pr(b|x)} \tag{3}$$

Equality ( 2 ) is a requirement of the neutral model resulting from the factorization ( 1 ), but expressed in terms of the observables $A$, $B$ and $E$ only.

## Abstraction from mechanistic details

The notation $E_A$, $E_B$, $E_Z$ and $Z$ enwraps the unobserved mechanistic structure in our logical framework. All these structures cancel in the derivation of the neutral model ( 2 )-( 3 ) (see Methods). Therefore, departure from ( 2 ) implies that for any resolution of the effect $E$ into a disjunction $E_A \vee E_B \vee E_Z$, requirement ( 1 ) cannot hold. In other words, there is no possible separation of the pathways leading to effect $E$ that can be independently associated to the factors $A$ and $B$. In this case information of the effects of independent exposure is not enough to predict their joint effect. Fortunately, because of this structure cancelation, searching for a particular splitting $E_A$, $E_B$ and $E_Z$ of the effect $E$ satisfying ( 1 ), that might not even exist, is not required for the diagnosis of interaction.

This is of crucial importance in practice. For example, large scale interaction studies are nowadays performed to build interconnected maps of simpler organisms(*12*). Millions of gene pairs are tested but only a very small fraction interacts. This already daunting task would be impossible if the molecular mechanisms mediating each pair were required *a priori* to detect interaction. Instead, the presence of interactions can be diagnosed by equality ( 2 ), and then subsequent experiments to discern interaction mechanisms can be specifically targeted in promising pairs, without wasting in the rejected pairs.

## Safety device against spurious susceptibility

If a factor not mechanistically associated to an effect is correlated to a causal factor, it can appear coincidentally associated to the effect, without actual causal connections. Such is the case of a "neutral" locus in the close proximity (linkage disequilibrium) to a locus causally connected to a given disease, mimicking the frequencies and correlations of the causal locus, displaying a spurious association to the disease.

Suppose $\Pr(E_B|b)$ is the risk of a mutation $b$ of a gene $B$ casually associated to cancer $E_B$. Let $A$ be a gene not causally associated to that cancer, such that $\Pr(E_B|x\,b) = \Pr(E_B|b)$ for every allele $x \in A$. If some selective phenomenon unrelated to the disease introduces structure in the distribution of genes, such that $\Pr(x\ b) \neq \Pr(x) \Pr(b)$, we have

$$\Pr(E_B|x) = \sum_{b \in B} \Pr(E_B|b) \Pr(b|x), \quad \forall x \in A \tag{4}$$

which associates gene $A$ to the effect $E_B$, even when the molecular machinery involved in the disease is not perturbed by this gene. The right side of ( 4 ) can be interpreted as the expected risk originated from the variants of causal factor $B$ in the proportions they co-occur with the variant $x$ of factor $A$. The structure $\Pr(x\ b)$ of the population spuriously "propagates" the susceptibility of factor $B$ to factor $A$, and that is why neutral genes are often erroneously associated to diseases.

The terms $\Pr(\overline{E}|x)$ and $\Pr(\overline{E}|y)$ introduce spurious susceptibilities $\Pr(\overline{E_B}|x)$ and $\Pr(\overline{E_A}|y)$ in the numerator of ( 3 ) (see also equation ( 10 ) in Methods). These fake associations cancel with the denominator, and the neutrality function ends up depending only on the pure susceptibility carrier $\Pr(\overline{E_A}|a)$ and $\Pr(\overline{E_B}|b)$ (see equation ( 11 )-( 12 ) in Methods).

### Invariance to population structure

Along with the cancellation of the propagated association also went away the dependency on the structure $\Pr(a\ b)$ of the joint distribution of factors variants shown in the denominator of ( 3 ) (See equation ( 12 ) in Methods). Hence, the neutrality function derived here is invariant to the marginal distribution of factors.

Correlated structure of $\Pr(a\ b)$ is ubiquitous in populations, controls, and patients, and their impact on inferences has been realized for long in population genomics and genetics. However, so far as we know, no such built-in cleaning device has been revealed in previous approaches to the interaction subject. This issue has been ignored, or patched *a posteriori* at best. The correlation cleaning can be particularly relevant for experimental design and data analysis in big-data scenarios, like cancer genome projects.

### Interaction measure

The background factor distribution $\Pr(a\ b)$ is required in the denominator of ( 3 ). However, the structure of $\Pr(a\ b)$ is often unknown, and gathering related information usually requires expensive experimental designs. One way to get rid of these factors is to take ratios.

$$\mathcal{J}_{ab} = \frac{\Pr(\overline{E}|a\ \overline{b}) \Pr(\overline{E}|\overline{a}\ b)}{\Pr(\overline{E}|\overline{a}\ \overline{b}) \Pr(\overline{E}|a\ b)} \tag{5}$$

Where $a, \overline{a} \in A$ and $b, \overline{b} \in B$. Note that $\mathcal{J}$ is expressed in terms of observables only, but without requiring the distribution $\Pr(a\ b)$ to be computed. Replacing each probability term in ( 5 ) by ( 1 ), reduce the non-interaction hypothesis to $\mathcal{J} = 1$, or by taking logarithms

$$H_0: \log \mathcal{J}_{ab} = 0 \tag{6}$$

The magnitude of interaction is an increasing function of the module of $\mathcal{J}$, and $\log \mathcal{J} > 0$ or $\log \mathcal{J} < 0$ indicates positive or negative interactions respectively (see demonstration in Methods, equations ( 16 )-( 20 )). Equations ( 5 ) and ( 6 ) lead to a model multiplicative in the complement of the effect,

$$\frac{\Pr(\overline{E}|a\ b)}{\Pr(\overline{E}|\overline{a}\ \overline{b})} = \frac{\Pr(\overline{E}|a\ \overline{b})}{\Pr(\overline{E}|\overline{a}\ \overline{b})} \frac{\Pr(\overline{E}|\overline{a}\ b)}{\Pr(\overline{E}|\overline{a}\ \overline{b})} \tag{7}$$

which can also be verified directly from ( 1 ).

Despite this model has been advocated in toxicology (*21*, *22*), it is mostly ignored by the dominant epidemiology and genetic networks literature(*1*, *13*, *23*).

## DETECTING GENETIC INTERACTIONS

### Interaction and fitness

Several genome-scale interaction studies have been conducted in *Saccharomyces Cerevisiae*(*13*), and the recent literature illustrates the quandary in the resolution of the concepts of interaction and fitness (see Methods). Can the measures used in these studies be explained in terms of the model multiplicative in the complement of the effect? How these models perform in the detection of interaction? To address these interrogations in a unified manner, the problem originally contextualized in terms of genes and fitness, is reformulated here in terms of factors and effect.

Let $\lambda$ denotes the average rate of cell duplication per unit of time. The probability for a strain $xy$ that cell duplicates at least once in a lapse of time $t$ (i.e. $E \equiv n > 0$) can be modeled according to (*24*) by

$$\Pr(E|xy) = 1 - e^{-\lambda_{xy} t} \qquad (8)$$

In this case $x$ and $y$ might denote gene variants at two locus $A$ and $B$. We choose, without losing generality, the duplication average time of wild-type strain as the unit of time, i.e. $\lambda_{00} = 1$. Substituting ( 8 ) in ( 5 ) and taking logarithm, the measure ( 6 ) for genetic interaction case yields

$$\log \mathcal{J}_{ab} = (1 + \lambda_{ab}) - (\lambda_{\overline{a}b} + \lambda_{a\overline{b}}) \qquad (9)$$

In experiments where cell grow at a constant rate, the parameter $\lambda$ is not the duplication rate but the growing rate, $n$ is not the number of duplications, but the number of cell. Derivations ( 8 )-( 9 ) follow faithfully.

By posing the genetic interaction problem in terms of probabilities of factors and effect we have shown that the multiplicative model in the complement of the effect imply additivity of duplication rates ( 9 ), demonstrating in the way, that the measure of interaction used by Janos et. al. can be justified from the basic principles appealed here.

In what follows, performance of the derived measure are evaluated with data from the largest global interaction network study in *Saccharomyces Cerevisiae*(*12*). We compare the measure ( 9 ), with the multiplicative on rate measure $\mathcal{M} = \lambda_{11} - \lambda_{01}\lambda_{10}$ used in the original study, where $\lambda$ are the growth rate relative to wild-type of single-mutants (01 and 10) and double-mutant (11) isogenic cell populations (see Data source in Methods for more details).

### Interaction networks

The overall impact on the interaction network is shown in the $\log \mathcal{J}$ vs. $\mathcal{M}$ plot in Figure 2A of the Essential x Essential (ExE) genes pairs dataset. A similar plot is obtained for Nonessential x Nonessential (NxN), see Figure S 1A. The initial correlation between the most extreme negative interactions (lower-left quadrant), is progressively deteriorated, due to a remarkable propensity of the measure $\log \mathcal{J}$ to score positive interactions with respect to $\mathcal{M}$.

We adopted from the original work(*12*) the regions $|\mathcal{M}| > 0.08$ to classify interaction at *p-value* < 0.05. The symmetric regions $|\log \mathcal{J}| > 0.0886$ contains the same number of interacting pairs at *p-value* < 0.05 for the ExE data. The overbar indicates no interaction, the symbol $\overline{M} J$ denote the number of pairs regarded as no-interaction by $\mathcal{M}$ and interaction by $\mathcal{J}$; the symbol $M \overline{J}$ those pairs regarded as interaction by $\mathcal{M}$ and no-interaction by $\mathcal{J}$; the symbol $MJ$ those pairs regarded as interaction by both $\mathcal{M}$ and $\mathcal{J}$; and the symbol $\overline{M}\ \overline{J}$ those pairs regarded as no-interaction by both $\mathcal{M}$ and $\mathcal{J}$.

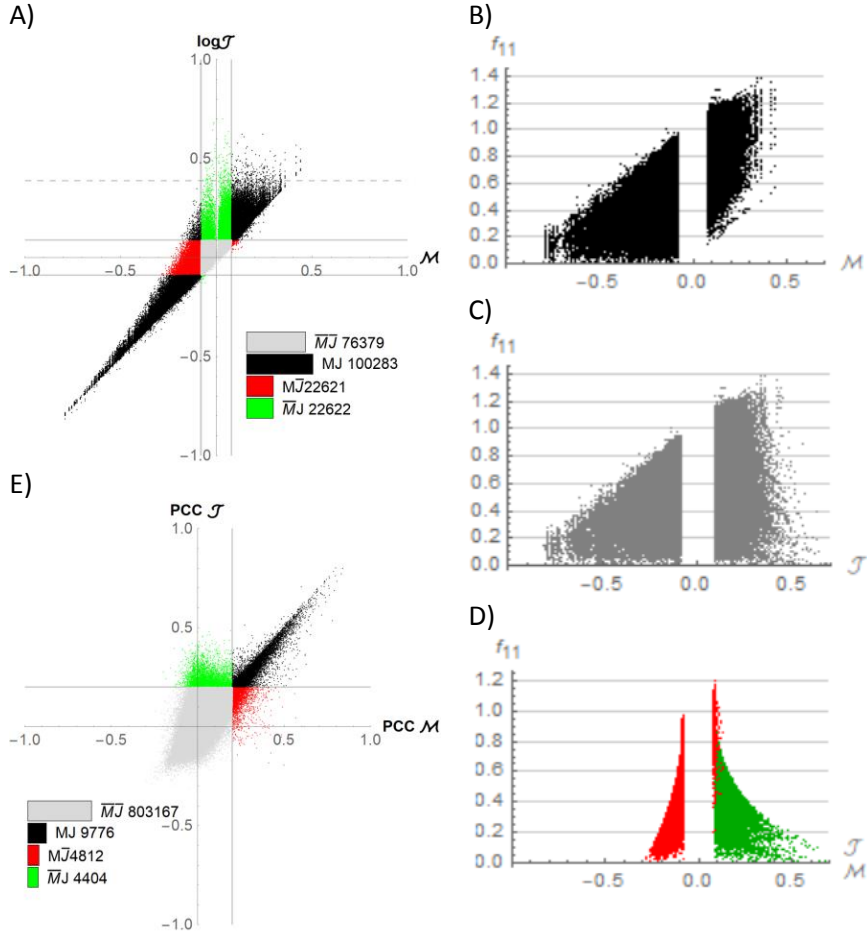

*Figure 2: J vs M differential scoring of interacting gene pairs retrieved from the Essential X Essential SGA library constructed in Saccharomyces Cerevisiae (12). Black dots indicate MJ co-identified interactions, red dots $M\overline{J}$, green dots $\overline{M}J$ and light-gray dots indicates $\overline{M}\ \overline{J}$. A) Comparison of the magnitude of interaction computed from the multiplicative measure $\mathcal{M}$ (X axis), and the additive measure $\log \mathcal{J}$ (Y axis). The log bar-chart inset represents the $\overline{M}J, M\overline{J}$, $MJ$ and $\overline{M}\ \overline{J}$ frequencies. Vertical and horizontal lines delimit the no-interaction region enclosed by -0.08 < $\mathcal{M}$ < 0.08 and -0.0886 < $\log \mathcal{J}$ < 0.0886. Dashed line indicate the over-conservative interaction threshold $\log \mathcal{J} > 0.34$ used to illustrates strong interactors in candidate hubs missed by $\mathcal{M}$. B) Double mutant fitness f11 vs M interaction score comprising the interacting gene pairs detected within the ExE library. C) like B) but f11 vs J. D) $M\overline{J}$ (red dots) and $\overline{M}J$ (green dots) with coordinates M and J respectively. E) Functional similarity (PCC). Vertical and horizontal lines at PCC=0.2 delimit the similarity threshold.*

About 18% of the interactions assigned by one measure are missed by the other. Hence, if one of the measure were correct, the other measure would miss 18% of the interactions, while 18% of the reported interactions are spurious, i.e. $(\overline{M}J/(\overline{M}J + MJ)$ or $M\overline{J}/(M\overline{J} + MJ)$ ). Thus, more than 30% of genetic interactions ( $(\overline{M}J + M\overline{J})/(\overline{M}J + M\overline{J} + MJ)$ ) are redefined by one or the other measure.

The relative frequency of negative genetic interactions detected by $\mathcal{M}$ exceeds the number of positive interactions in both ExE and NxN libraries (Figure S 1B, and Figure 1A in (12)). However, by using $\mathcal{J}$ the pattern reverses with positive interactions now prevailing in the ExE library. Within this library, $\mathcal{J}$ scored more than 20 000 new positive interactions, whereas the number of negative interactions decreased to a similar extent (Figure 2A). An even distribution of positive and negative interactions is produced by $J$ with the NxN library (Figure S 1B).

Major discrepancies between $J$ and $M$ in detecting positive genetic interactions arise for double mutant fitness values below 0.5, , spanning interaction J-scores from 0.2 to 0.7 in both libraries (empty triangle in the positive side of Figure 2B, filled by $J$ in Figure 2C, and similar in Figure S 1C dark-gray dots). This suggests that $M$ measure would be particularly insensitive to low double-mutant fitness values. On the other hand, negative genetic interactions are consistently captured by both measures across different magnitudes of interactions and fitness (lower-left quadrant in Figure 2A, and left side of Figure 2B-C, Figure S 1A-C).

### Functional Annotations

To evaluate how the interacting gene pairs segregates into functional categories with each measures we computed the number $\overline{M}J$, $M\overline{J}$ and $MJ$ of pairs co-annotated to the same biological process (GO BP), pathway (KEEG), or molecular complexes (Data S 1-Data S 3). On average, $\mathcal{M}$ missed 31% to 44% of the positive interactions annotated by $\mathcal{J}$, more notable within molecular complexes (Figure S 2A); whereas positive interactions missed by $\mathcal{J}$ did not exceeded 6%. On the other hand, $\mathcal{J}$ missed 1.6% to 25% of the negative interactions annotated by $\mathcal{M}$ along the three functional categories. A similar global pattern was obtained from the interaction analysis of the NxN library (Data S 5-Data S 7; Figure S 3A).

Subsequently, we highlighted processes, pathways and molecular complexes with higher rates of annotated interactions missed by either measures within the ExE library (Figure S 2B-C; Figure S 4) and the NxN library (Figure S 3B-G).

#### Biological Processes

More than 25% of the positive interactions reported by J, spanning 140 of 189 GO_BP terms with 500 or more interactions, are not reported by $M$ (Data S 1). Particularly, the top percent (> 50%) pertained to "ribonucleoprotein complex subunit organization" and "ribonucleoprotein complex assembly", where J detected 567 and 545 positive interactions ($MJ + \overline{M}J$) respectively; while $M$ only detected 267 and 257 of them. Comparatively, J missed 0.5% to 13% of the positive interactions reported by $M$.

On the other hand, 5% to 36% of the negative interactions reported by $\mathcal{M}$ are missed by $J$. In 71 of 234 GO_BP categories, the number of interactions missed by $\mathcal{J}$ (i.e. $M\overline{J}$) represents more than 25% of the total number of interactions this measure report ($MJ + \overline{M}J$). Among the top 10 GO_BP categories with higher miss annotation rates for both positive and negative interactions, five are related to ribosome biogenesis (Figure S 2B).

#### Pathways

In 20 of the 35 pathways comprising more than 10 interactions, $M$ measure would miss between 25% and 100% of the positive genetic interactions scored by J (Data S 2, Figure S 2A, Figure S 4A-B). Of note, $\mathcal{M}$ missed more than 40% interactions reported by $\mathcal{J}$ in the pathways "Basal transcription factors", "Ribosome biogenesis in eukaryotes" and "Proteasome", summarizing more than 300 genetic interactions. Particularly, positively interacting gene pairs co-annotated to the pathway "Ribosome biogenesis in eukaryotes" emerged under-annotated by $M$.

Otherwise, negative genetic interactions were almost equally scored by both measures (Figure S 4B). The largest divergence among the negative interactions was found within the pathway "Ribosome biogenesis in eukaryotes" with 26 of 243 interacting pairs classified by $M$ as negative interacting pairs (10.7%). Altogether, roughly 120 genetic interactions pairs co-annotated to the pathway "ribosome Biogenesis in eukaryotes" are differentially scored by $M$ and J (56.9%).

Molecular complexes

Among the molecular complexes defined by Costanzo et al. 2016 containing more than 10 interaction pairs, 30 and 54 comprised positive and negative interactions respectively (Data S 3). Overall, negative interactions were roughly 15-fold higher than positive interactions across the molecular complexes. Two major and related complexes, the "preribosome-large subunit precursor" and "90S preribosome" comprised a large number of divergent annotations between $J$ and $M$ (Figure S 2C). More than 50% of positive genetic interactions are missed by $M$, whereas 30-40% of negative interactions are missed by $J$.

Missing hubs

J and M agree on more than 80% of the detected interactions, and several hubs are co-identified within ExE and NxN libraries (Data S 8). Both measures produce hubs with similar degree, identity and magnitude of the genetic interactions (Figure S 5A-B left). However, the spread above the diagonal in the plots of connections degree suggest the detection of highly connected genes by $J$, not detected by $M$ in both libraries (Figure S 5A-B right).

Table 1 Candidate hubs obtained with the J measure that are missed by the M measure ($\overline{M}\ J$) identified from ExE and NxN libraries.

| candidate hubs | Library | $\overline{M}\ J$ | $M\ \overline{J}$ | $M\ J$ | $M$ | $J$ |
|---|---|---|---|---|---|---|
| **trm112** | ExE | 169 | 5 | 2 | 7 | 171 |
| **tif35** | ExE | 164 | 6 | 2 | 8 | 166 |
| **noc4** | ExE | 163 | 1 | 4 | 5 | 167 |
| **rrp7** | ExE | 117 | 2 | 4 | 6 | 121 |
| **tim17** | ExE | 97 | 1 | 2 | 3 | 99 |
| **ZAP1** | NxN | 138 | 0 | 2 | 2 | 140 |
| **vma7** | NxN | 131 | 1 | 3 | 4 | 134 |
| **msm1** | NxN | 110 | 1 | 5 | 6 | 115 |
| **rpb4** | NxN | 117 | 6 | 2 | 8 | 119 |

Thus, we explore genes which according to $M$ has less than 10% of the number of interactions only captured by J, i.e. satisfying $M\ J + M\ \overline{J} < 0.1 \overline{M}\ J$. The candidate hubs so captured from the ExE and NxN libraries are listed in Table 1 and the distribution and magnitude of the interactions can be appreciated from Figure 3A and B for both measures. Notably, the few connections detected only by $M$ lay in the borderline close to $M = -0.08$.

The five candidate hubs surfaced by $J$ within the ExE library are annotated with 27 to 81 physical interactions, and 10 to 41 genetic interactions (SGD; http://www.yeastgenome.org/), and their Temperature Sensitive (TS) alleles significantly decrease the fitness (between 0.2037 and 0.3420) as expected for a hub protein (Table S1). Moreover, the five genes are pleiotropic as verified by the multiple Gene Ontology annotations using GO-Slim terms (*25*).

The candidate hubs tif35 and tim17 are essential components of molecular complexes partaking protein translation and mitochondrial import channel structure (*26*, *27*). The genes trm112, noc4 and rrp7 are involved in ribosome biogenesis and export, and located in the nuclear compartment of the cell (*28–30*). Accordingly, these genes displayed expression correlation with a set of 20 genes enriched for the GO_BP ribosome biogenesis (Data S 4: SPELL analysis ACS>5.3, p-value=7.31e-23)(*31*). Moreover, trm112, noc4 and rrp7 are included in the GO_BP categories "ncRNA processing" and "ribonucleoprotein complex biogenesis" which comprised a large number of interactions missed by $M$ (Figure S 2B; Data S 1). Noc4 is also a member of the protein complex "90S preribosome", a category where $M$ missed about 70% of the positive interactions reported by $J$ (Figure S 2C; Data S 3). Altogether, the five candidate hubs displayed more than 700 previously unnoticed positive interactions, spanning at least three different biological processes and two cellular compartments (Table 1 and Table S1).

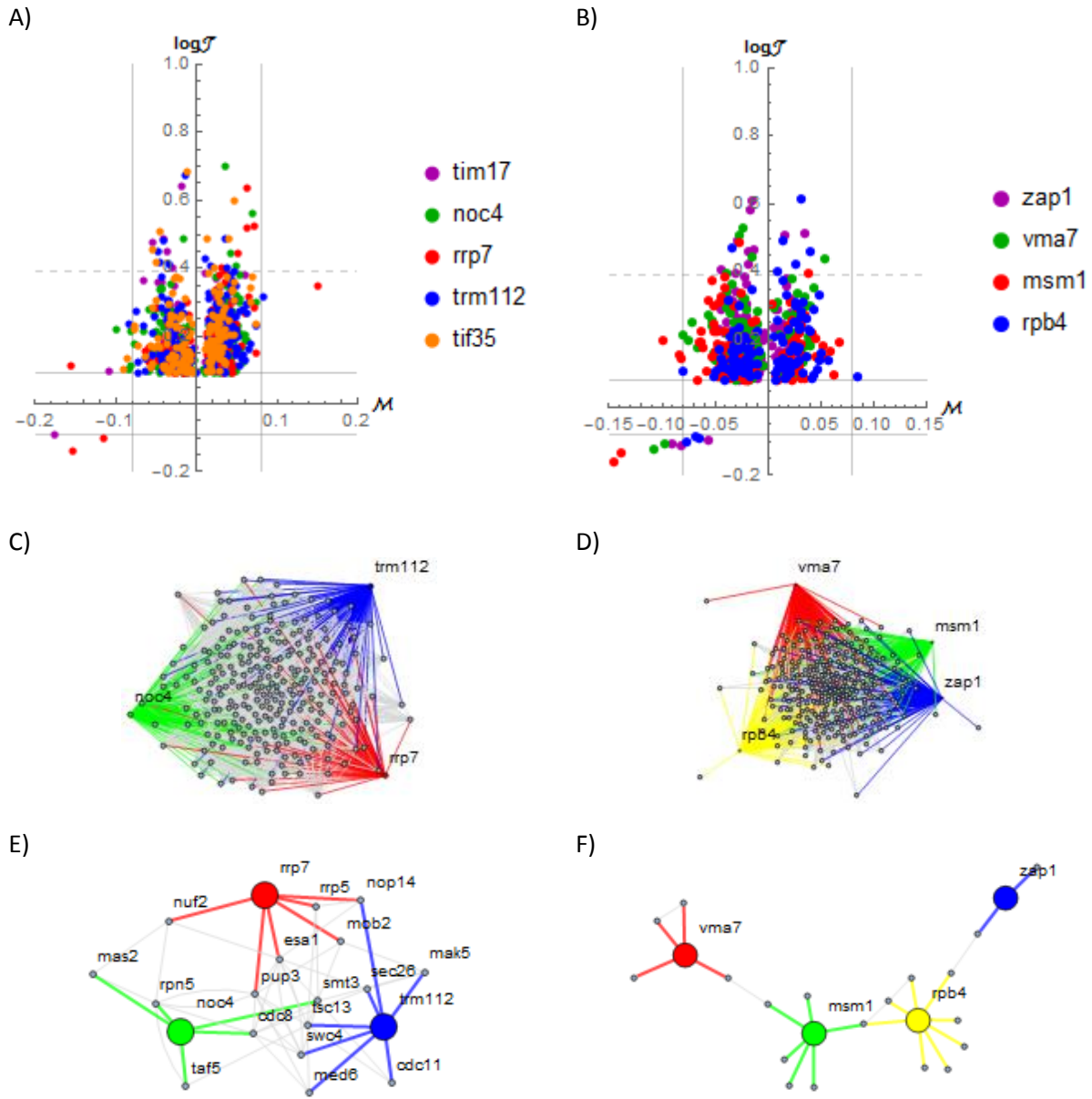

Figure 3: Candidate hub missed by M listed in Table 1. A) M vs. J of the ExE library. B) M vs. J of the NxN library. C) Trm112, noc4 and rrp7 interacting networks according to J. E) according to M. D) zap1, vma7, msm1, rpb4 interacting networks according to J. F) according to M.

The extensive re-wiring of the local genetic interaction network of the hubs trm112, noc4 and rrp7 is illustrated in Figure 3C-E. Although no direct genetic interaction was detected among them, $J$ identified 123 intermediary connectors (distance 1). Notably, these set of connecting genes were enriched for Biological Processes such as RNA processing (40/123 genes, p<1.34e-6), ribonucleoprotein complex subunit organization (21/123, p<1.47e-5) and ribonucleoprotein complex biogenesis (34/123, p<1.47e-4) (see Method). In contrast, $M$ found nop14 as the unique intermediary connector among two of these hubs, and the interaction network is pretty sparse (Figure 3E).

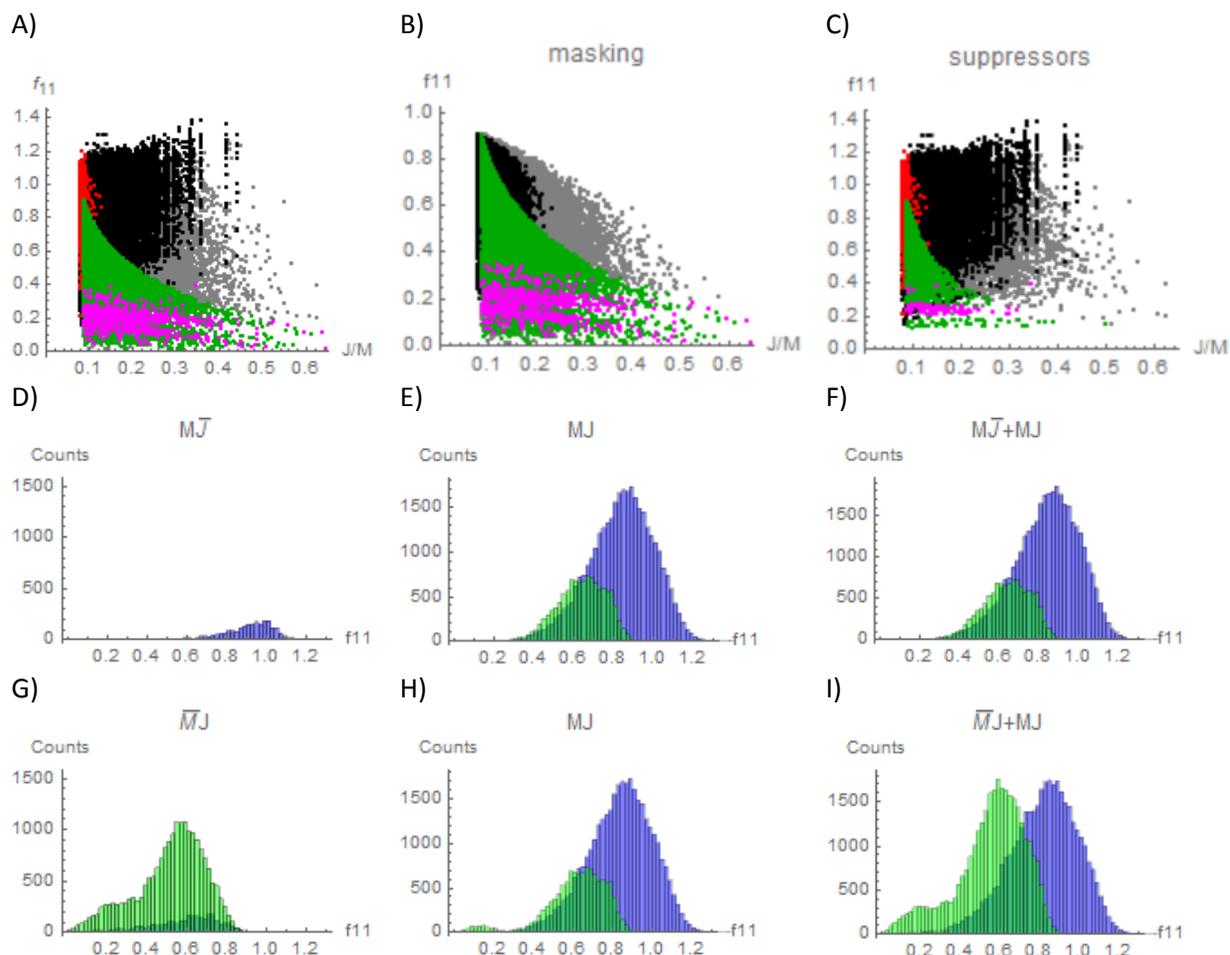

*Figure 4: Double mutant fitness (f11) vs J/M interaction score comprising positive interactions (ExE library). Black and dark-gray dots indicate MJ co-identified interactions the first with coordinates M and the later with coordinate J. Red and green dots indicates $M\overline{J}$ and $\overline{M}J$ interactions. The newly identified hubs are highlighted in magenta. A) All positive interactions. B) masking interactions. C) Suppressor interactions. D-F) distribution of double mutant fitness f11 of the interaction detected by M. G-I) distribution of double mutant fitness f11 of the interaction detected by J.*

Four candidate hubs surfaced from the NxN library after a similar analysis, summarizing more than 490 new positive interactions (Table 1). These hubs are annotated with several physical and genetic interactions and their deletion in the yeast genome significantly decreased cell fitness (Table S 1). Particularly, the candidate hubs rpb4 and zap1 are involved in global (rpb4) or more specific (zap1) gene expression programs including transcription (rpb4, zap1), mRNA-decay and export (rpb4), and translation initiation (rpb4) (*32, 33*); likewise they are annotated in two or more GO-Slim terms (Table S 1). Relevant biological processes where rpb4 and zap1 partake, including "translation", "RNA processing" and "gene expression", are notably missed by $M$ (>35%)(Figure S 3B). The hubs zap1, vma7, msm1 and rpb4, surfaced by $J$, and their derived genetic interactions remarkably changed the local topology of the interaction network wired by $M$ (Figure 3D-F).

A symmetric analysis was preformed to detect $M\overline{J}$ candidate hubs with more than 10 connections only detected by $M$ (i.e. $M\overline{J} > 10$), where J missed more than 90% (i.e. $MJ + \overline{M}J < 0.1M\overline{J}$). No candidates were gathered in any of the libraries (ExE and NxN), neither even with a less restrictive filter $MJ + \overline{M} < M\overline{J}$, suggesting that J measure is sensitive to detect meaningful interactions out of the reach of M.

### Masked and suppressor interactions

Considering that major divergences between the interactions measures comprise positive genetic interactions surfaced only by J, we further classify them into suppressor, $f_{11} > \min(f_{01}, f_{10})$, or masking interactions (the reminder positive interaccions). Moreover, we analyzed the distribution of the emerging hubs listed in Table 1 across these types of alleviating genetic interactions (Figure 4 and Figure S 7). Most of the newly detected positive interactions comprised masking interactions (19703 of 22622, 87%), resulting in double mutant (f11) fitness values within 0.1-0.8 (Figure 4G-I). On the contrary, masking interactions producing f11<0.4 are poorly detected by M (Figure 4D-F). The amount of J-surfaced masking interactions equates the number of suppressor interactions detected in the ExE library (Figure 4I).

On the other hand, the five ExE hubs partake 643 masking (89%) and 81 (11%) suppressor interactions. Such hubs are surfaced by J at fitness values f11<0.4 (Figure 4B-C). Particularly, trm112, noc4 and rrp7 which are involved in Ribosome Biogenesis displayed 78 positive interactions within this process, 70 masking and eight suppressors.

# DISCUSSION

The resolution of the concept of interaction accomplished from first principles here, recognize the inclusion of multistage distant interactions as a more realistic approach. The implied neutrality function lead to the model of no interaction multiplicative in the complement of the effect. The derivations revealed theoretical properties unnoticed so far, that turned out particularly relevant for genetic interaction mapping in large scale assays, where distribution of gene variants differ between populations, and genetic susceptibilities are spuriously propagated by linkage disequilibrium. The warranted abstraction from the intricacies of molecular mechanisms mediating the interactions, is of crucial practical importance for global genetic interaction mapping, and other big-data scenarios.

The choice of the interaction measure determines the identity, sign, strength and distribution of genetic interactions across functional modules and subcellular compartments of the cells, conditioning the functional mapping. Genetic interaction profiles has been used to assemble hierarchical models of cell function on Saccharomyces Cerevisiae (*12*). The gene-fitness interaction question, reformulated in terms of factors and effect, lead to a simple additive formula also benefited from the theoretical niceties here revealed, lacking in other interaction measures. The comparative performance shown on the genetic interaction data imply:

- Partial re-wiring of regions of the currently known genetic interaction landscape in yeast with functional implications and relevancy to comparative genomics studies.
- Less aversion to positive interactions, re-assessing their roles in biological processes, pathways, and molecular complexes, now suggesting that positive interactions prevail among essential genes.
- Capturing experimentally supported hubs mostly enriched by positive interactions, otherwise missed by current multiplicative measure.
- That masking interactions have a more important role on ribosome biogenesis than previously conceived.

It has been previously considered that interaction models become shady with low fitness double mutants(*15*). Precisely, major divergence between M and J comprise positive genetic interactions of the masking type resulting in low fitness phenotypes (<0.3) and spanning a wide range of interaction scores (0.2<J<0.7).

Costanzo et. al. gauge the functional similarity between genes with the Pearson correlation coefficient (PCC) of the genetic interaction profiles, and create similarity networks that organize genes into clusters highlighting biological modules(*12*). Figure 2E emphasizes the discrepancies between PCC computed with J and M measures (ExE library). Of note, more than 30% of pairs reported with similarity PCC > 0.2 by one measure is missed by the other. Since the genetic interaction profile of a particular gene is composed of its specific array of negative and positive interactions, it is expected that dissimilar annotation of such interactions redefine some functional domains, as well as modify other profile similarity-based derivations like prediction of gene function, pleiotropy and network connectivity (*12*).

The architectural features of complex systems, found in social and technological networks are shared by molecular interaction networks within a cell. This universality suggest that similar laws govern most complex networks in nature. Nothing in the derivation of the interaction model restrict the scope to network biology or biomedical applications. The unity of arguments elaborated here permit to analyze dissimilar interaction models and experimental data within a common framework. We hope this unified view contributes towards the end of a vivid controversy and complaisance with the coexistence of different model of interaction that plague the literature with inconsistent revenues.

Systematic analysis of tripartite or more complex genetic interactions networks commence to emerge(*34*)(*35*). Hence, theoretical research and experimental designs remain critical for more accurate construction of higher-order interactions landscape.

# MATERIALS AND METHODS

## Neutrality function derivation

Multiplying both sides of $\Pr(\overline{E}|x\,y) = \Pr(\overline{E}_A|x)\Pr(\overline{E}_B|y)\Pr(\overline{E}_Z)$ in ( 1 ) by $\Pr(b|x)$, and summing over $b \in B$, and doing the same but with $\Pr(a|y)$ over $a \in A$, yields

$$\Pr(\overline{E}|x) = \Pr(\overline{E}_A|x)\Pr(\overline{E}_B|x)\Pr(\overline{E}_Z)$$
$$\Pr(\overline{E}|y) = \Pr(\overline{E}_A|y)\Pr(\overline{E}_B|y)\Pr(\overline{E}_Z) \qquad (10)$$

Where $\Pr(\overline{E}_B|x) = \sum_{b \in B}\Pr(\overline{E}_B|b)\Pr(b|x)$ and $\Pr(\overline{E}_A|y) = \sum_{a \in A}\Pr(\overline{E}_A|a)\Pr(a|y)$. The factorization ( 1 ) can be written as

$$\Pr(\overline{E}|x\,y) = \frac{\Pr(\overline{E}|x)\Pr(\overline{E}|y)}{\Pr(\overline{E}_A|y)\Pr(\overline{E}_B|x)\Pr(\overline{E}_Z)} \qquad (11)$$

Expanding the product in the denominator and using ( 1 ) yields

$$\Pr(\overline{E}_A|y)\Pr(\overline{E}_B|x)\Pr(\overline{E}_Z)$$
$$= \Pr(\overline{E}_Z)\left\{\sum_a \Pr(\overline{E}_A|a)\Pr(a|y)\right\}\left\{\sum_b \Pr(\overline{E}_B|b)\Pr(b|x)\right\}$$
$$= \sum_{a \in A, b \in B} \Pr(\overline{E}_A|a)\Pr(\overline{E}_B|b)\Pr(\overline{E}_Z)\Pr(a|y)\Pr(b|x) \qquad (12)$$
$$= \sum_{a \in A, b \in B} \Pr(\overline{E}|a\,b)\Pr(a|y)\Pr(b|x)$$

Substituting ( 10 ) and ( 12 ) in ( 11 ) yields that for non-interacting factors

$$\Pr(\overline{E}|x\,y) = \frac{\Pr(\overline{E}|x)\Pr(\overline{E}|by)}{\sum_{a \in A, b \in B} \Pr(\overline{E}|a\,b)\Pr(a|y)\Pr(b|x)} \qquad (13)$$

fully expressed in terms of observables.

## Log-linearity of the neutral model

The logarithm of $\Pr(\overline{E}|x\,y)$ in non-interaction or interaction scenarios can be expressed in the form

$$\log\{\Pr(\overline{E}|x\,y)\} = \mu + \alpha_x + \beta_y - \delta_{xy}, \qquad \forall x \in A, y \in B \qquad (14)$$

where a generally valid assignment can be conveniently obtained by fixing

$$\mu = \log \Pr(\overline{E}|\bar{a}\,\bar{b})$$
$$\alpha_{\bar{a}} = \beta_{\bar{b}} = \delta_{x\bar{b}} = \delta_{\bar{a}y} = 0 \qquad (15)$$

Which imply

$$\alpha_a = \log\frac{\Pr(\overline{E}|a\,\bar{b})}{\Pr(\overline{E}|\bar{a}\,\bar{b})}, \qquad \beta_b = \log\frac{\Pr(\overline{E}|\bar{a}\,b)}{\Pr(\overline{E}|\bar{a}\,\bar{b})} \qquad (16)$$

$$\delta_{ab} = \alpha_a + \beta_b - \log \frac{\Pr(\overline{E}|a\ b)}{\Pr(\overline{E}|\bar{a}\ \bar{b})} \tag{17}$$

In the particular non-interaction case, the minimal requirement ( 1 ) implies that

$$\mu = \log \Pr(\overline{E}_A|\bar{a}) + \log \Pr(\overline{E}_B|\bar{b}) + \log \Pr(\overline{E}_Z)$$

$$\alpha_x = \log \Pr(\overline{E}_A|x) - \log \Pr(\overline{E}_A|\bar{a})$$

$$\beta_y = \log \Pr(\overline{E}_B|y) - \log \Pr(\overline{E}_B|\bar{b}) \tag{18}$$

$$\delta_{ab} = 0$$

Conversely suppose the logarithm of $\Pr(\overline{E}|x\ y)$ can be expressed in the linear form $\mu + \alpha_x + \beta_y$, where $\mu$ is a constant, $\alpha_x$ depends only on $x \in A$, and $\beta_y$ depends only on $y \in B$. The numerator in ( 13 ) becomes

$$\Pr(\overline{E}|x) = \sum_{b \in B} \Pr(\overline{E}|x\ b) \Pr(b|x) = \exp(\mu + \alpha_x) \sum_{b \in B} \exp(\beta_b) \Pr(b|x)$$

$$\Pr(\overline{E}|y) = \sum_{a \in A} \Pr(\overline{E}|a\ y) \Pr(a|y) = \exp(\mu + \beta_y) \sum_{a \in A} \exp(\alpha_a) \Pr(a|y)$$

The denominator in ( 13 ) becomes

$$\sum_{a \in A, b \in B} \exp(\mu + \alpha_a + \beta_b) \Pr(a|y) \Pr(b|x) = \exp(\mu) \sum_{a \in A} \exp(\alpha_a) \Pr(a|y) \sum_{b \in B} \exp(\beta_b) \Pr(b|x)$$

The summation terms cancel in the numerator and denominator of ( 13 ), yielding the condition ( 2 ) of no-interaction.

$$\mathcal{N}(x, y) = \exp(\mu + \alpha_x + \beta_y) \tag{19}$$

Therefore, except for normalization, a log-linear function is a neutral model. Replacing ( 14 ) in ( 5 ) yields that in general

$$\delta_{ab} = \log \mathcal{J} \tag{20}$$

indicating that $\mathcal{J}$ is a measure of the size of interaction. Further, positive or negative interaction arise when the joint effect is greater or smaller than expected, i.e. $\Pr(E|a\ b) > 1 - \mathcal{N}(a, b)$, or $\Pr(E|a\ b) < 1 - \mathcal{N}(a, b)$, or equivalently when $\delta_{ab} > 0$ or $\delta_{ab} < 0$, as can be corroborated from ( 14 ).

### Meaning of log-linear parameters

Far from merely being a convenient mathematical approximation valid within a limited subset of the neutral functions, the log-linear form of the neutral functions is a general valid model for no-interaction (*21*).

Suitable meanings of the parameters $\mu$, $\alpha_x$, $\beta_y$ can be suggested from the probability terms in ( 16 ) and ( 18 ), explicitly relating unobserved structure of inner mechanism to observed ones. In the particular non-interaction scenario, according to the neutral function ( 19 ), the exponential of $\alpha_x + \beta_y$ is the probability of $\overline{E}$ relative to no exposition.

$$\frac{\Pr(\bar{E}|x\,y)}{\Pr(\bar{E}|\bar{a}\,\bar{b})} = \exp\{\alpha_x + \beta_y\} = \exp\{\alpha_x\}\exp\{\beta_y\} \qquad (21)$$

The multiplicative model on the complement of the effect, equality XXX, is obtained from ( 16 ) and ( 21 ).

### Genes and fitness

We inspect various interaction networks studies in *Saccharomyces Cerevisiae* from the recent literature that collected growth measurements of wild-type (00), single-mutants (01 and 10), and double-mutant (11) isogenic cell populations (*13*). Even when they addressed the same interaction question, and advocated the same null multiplicative model $f_{11}f_{00} = f_{01}f_{10}$ on fitness $f$, their definitions of fitness differ, and so are their predictions(*13*).

Jasnos et. al. assayed growth curves of the resulting progeny of 639 randomly crossed pairs of isogenic individuals with deletions performing slow growth rates in one of 758 genes (*11*). These authors defined fitness $f$ by the factor $e^\lambda$, of a population growing continuously at a rate $\lambda$, and chose the null model as the log-fitness scale $\epsilon = (\lambda_{00} + \lambda_{11}) - (\lambda_{01} + \lambda_{10})$, which become additive on rates.

Onge et. al. studied the interaction of 650 double-deletion strains, corresponding to pairings of 26 non-essential genes that confer resistance to the DNA-damaging agent methanesulfonate (MMS). These authors defined fitness of each deletion strain directly by its duplication rate $\lambda$, relative to that of wild type (*10*), i.e. $f = \lambda$. The null model take the form $\epsilon = \lambda_{11} - \lambda_{01}\lambda_{10}$, where $\lambda_{00} = 1$.

Costanzo et. al. wired the most extensive global genetic interaction network in *Saccharomyces Cerevisiae*, with over 23 million double mutants involving 5 416 different genes(*12*), including the first large-scale interaction network comprising ~120 000 pairs of essential genes. These authors modeled colony growth rate $\lambda$ based on the empirical observations that colony area $c$ scaled linearly with time (*36*, *37*), i.e. $\lambda\,t = c$, and defined fitness proportional to the rate of change of colony area relative to that of wild type, i.e. $f \propto \lambda$. Like Onge et. al., the null model take the form $\epsilon = \lambda_{11} - \lambda_{01}\lambda_{10}$, though their $\lambda$ means growth rate instead of duplication rate.

### Data source

We downloaded from http://thecellmap.org/costanzo2016/ the normalized interaction data files SGA_ExE and SGA_NxN(*12*). The normalization removed systematic biases in colony size arising from experimental factors, and a model of fitness and genetic interactions for each double mutant were fit to the normalized colony sizes(*12*). For our purpose, entries with NaN in any of the numerical fields or with negative fitness values were ignored. The columns named "Query single mutant fitness (SMF)", "Array SMF", and "Double mutant fitness" are here denoted $\lambda_{01}$, $\lambda_{10}$ and $\lambda_{11}$ respectively.

For the sake of fairness, we do not introduce further data processing. The plots and analysis are purposely maintained in a factual level, as close as possible to the normalized data of the original source. In doing so, the comparison accommodates well to the exposition of the original work.

### Functional Annotations

#### Biological Processes

The gene ontology (GO) biological process data and the yeast gene association files were downloaded from http://geneontology.org/ on August 19, 2014 (Ashburner et al., 2000). We considered only those BP with more than 500 interaction pairs scored by both measures without regarding interaction sign (TP). Then, we compute the fraction of gene pairs which displayed positive interaction according to J but not by M, and the fraction of gene pairs which displayed negative interaction according to M but not by J. (Data S 1)

Pathways

We considered *S. Cerevisiae* pathways annotated in KEGG with more than 10 interaction pairs scored by both measures without regarding interaction sign (TP). Only 37 of 50 pathways satisfied this condition (Data S 2).

Molecular complexes

We considered the 143 molecular complexes defined by Costanzo et al. 2016 with more than 10 interaction pairs scored by both measures without regarding interaction sign (TP). Only 54 complex satisfied this condition (Data S 3).

Expression profile similarity

Genes with expression profiles most similar to the query gene set rrp7, noc4 and trm112 were detected by using the query-driven search engine SPELL Version 2.0.3 for Yeast at ip-172-31-19-98 with 512 datasets on 2018-07-16.

Gene Ontology Enrichment

GO terms enriched in the interaction network generated by J were gathered by the Gene Ontology Enrichment tool from the YeastMine website (Balakrishnan et al., 2012). The following parameters were selected: Ontology: Biological Processes; Test Correction: Holm-Bonferroni; P value: 0.05.

**Shared Hubs**

We select genes with more than 100 common connections and less than 5% of discordant connections, i.e. $\overline{M} J + M \overline{J} < 0.05 \, M J$, for $M J > 99$. With this hasty criteria, 16 and 339 hubs were identified within the ExE and NxN libraries, respectively. For instance, the hub mcm3 comprised 129 ExE common interactions (64 positive and 65 negative), with only one interaction scored exclusively by M (Data S 8). The genetic interaction network comprising such common hubs is only marginally affected by the interaction measure (Figure S 6).

# ACKNOWLEDGMENTS

## Funding
This work has been partially supported by the Center for Genetic Engineering and Biotechnology.

## Author contributions
J.F.C. Conceived and formulated the interaction problem into a probabilistic framework, made the mathematical derivations, and reformulated in terms of gene-fitness interactions. Drafted the paper.

J.F.C.D Conceived the factorization of the null model of interaction. Contributed and checked the mathematical derivations, and reformulated in terms of gene-fitness interactions.

Y.P. Propose the study case data, instantiate the biological questions and interpretations.

All authors participated in the interpretation of the formulas, in the analyses of genetic interaction data and results, conceived plots and figures, and revised and edited the paper.

## Competing interests
Authors declare no competing interests

## Data and materials availability
The source and availability of the data is described in the main text.

# SUPPLEMENTARY INFORMATION
# DETECTION OF GENETIC INTERACTIONS FROM FIRST PRINCIPLES

J Fernandez-de-Cossio, J Fernandez-de-Cossio-Diaz, Y Perera.
Correspondence to jorge.cossio@cigb.edu.cu

FIGURES S1 TO S8

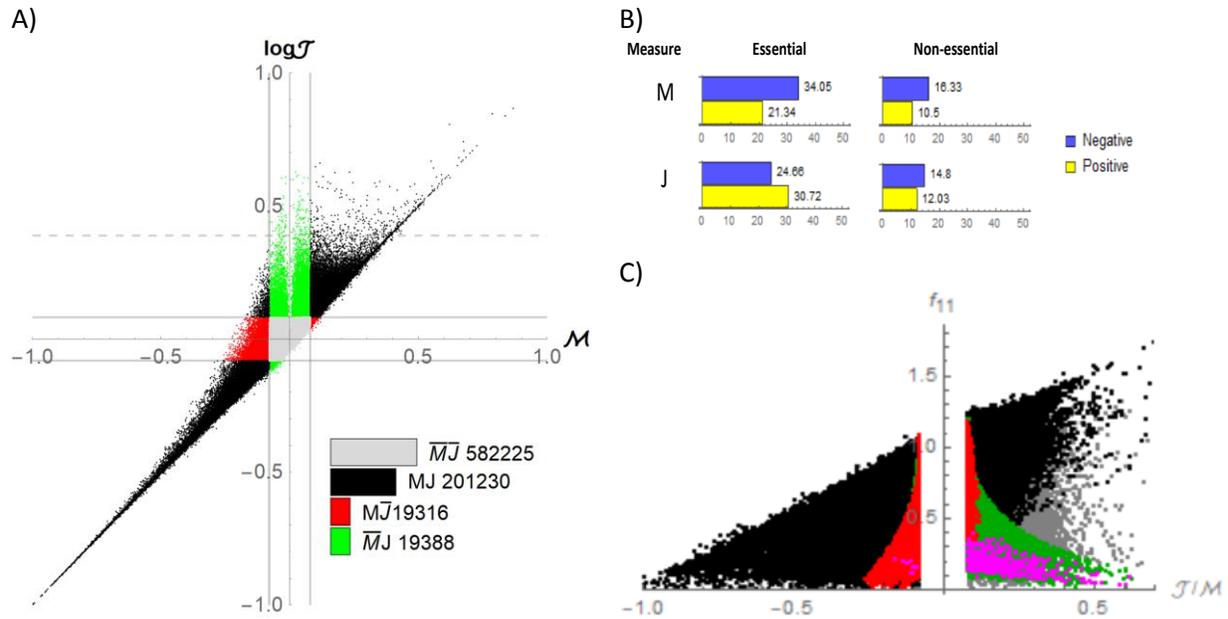

*Figure S 1: A) J vs. M plot of Nonessential x Nonessential gene pairs. B) Percent of positive and negative genetic interactions detected by $\mathcal{M}$ and $\mathcal{J}$ within the ExE and NxN libraries. Only those interactions with p<0.05 were included C) Double mutant fitness f11 vs J/M interaction score plot comprising the interacting gene pairs detected within the ExE library. $M\overline{J}$ (red dots) and M J (black dots) are plotted with coordinate M. $\overline{M}$ J (green dots) and M J (dark-gray dots) are plotted with coordinate J. newly detected hubs are plotted by pink dots.*

A)

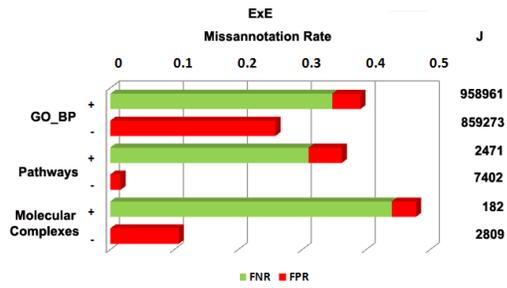

B)

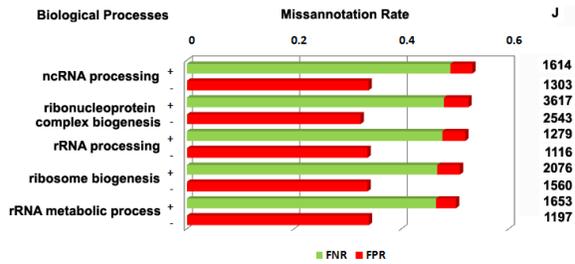

C)

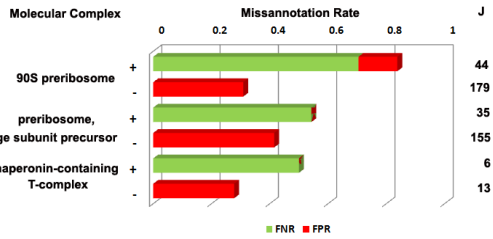

*Figure S 2: Discordant functional annotation of interacting gene pairs across Biological processes, Pathways and Molecular Complexes. The positive and negative interactions are separated in each category, and the $\overline{M}J$ and $M\overline{J}$ rates are stacked within each bar colored in green and red respectively. B,C) Top rated $\overline{M}J$ and $M\overline{J}$ annotation from the ExE library in Biological Processes (B) and Molecular Complexes (C).*

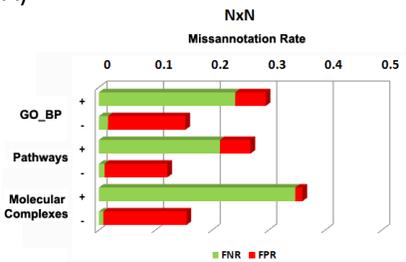
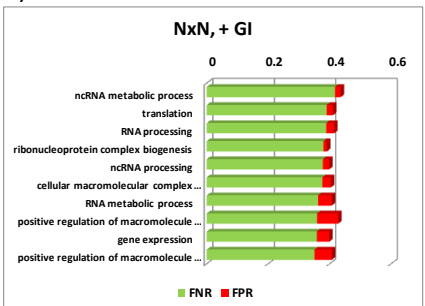
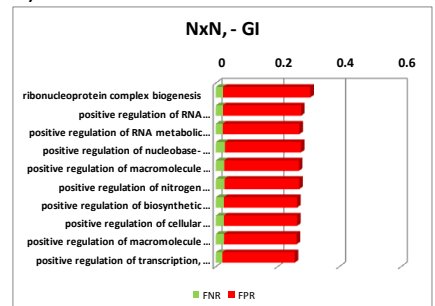
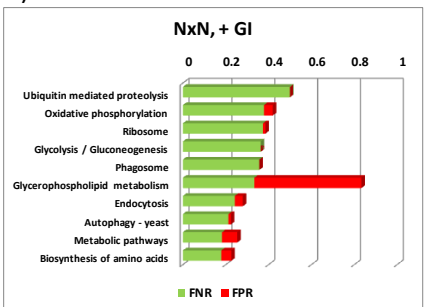
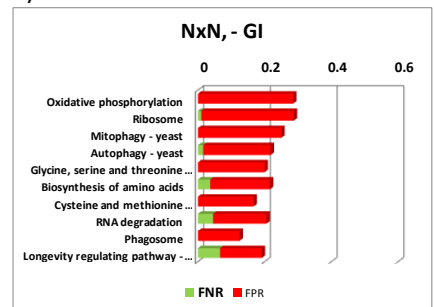
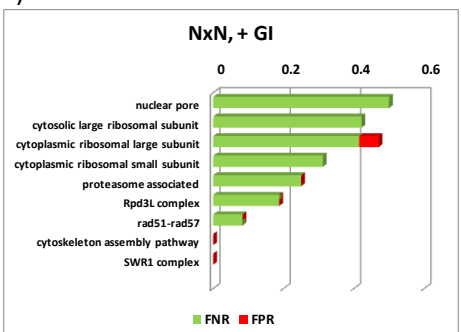
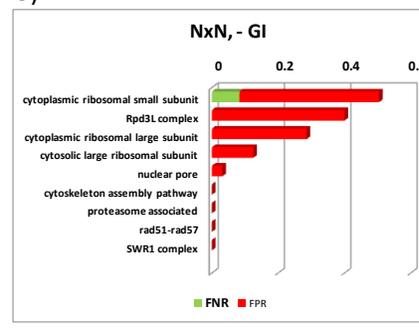

*Figure S 3: A) Discordant annotation of interacting gene pairs across Biological processes, Pathways and Molecular Complexes within the NxN library. The positive and negative interactions are separated in each category, and the FN and FP rates are stacked within each bar colored in green and red respectively. B-G) Top ten miss annotated Biological Processes, Pathways and Molecular Complexes for Positive and Negative genetic interactions.*

A) 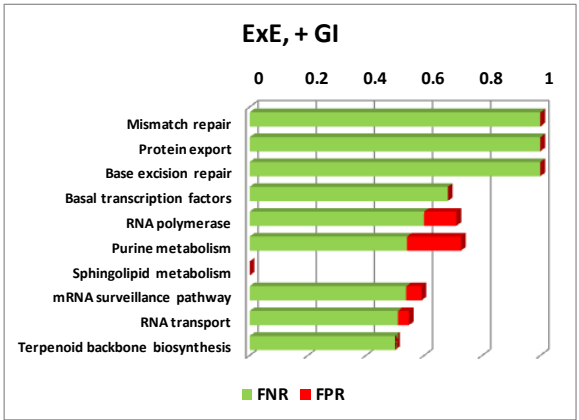
B) 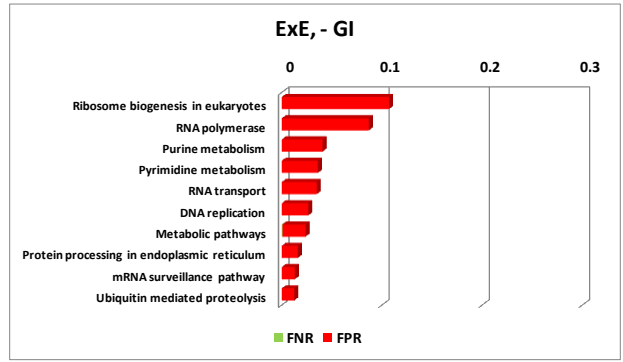

*Figure S 4: Top ten miss annotated pathways according to Positive and Negative genetic interactions detected within ExE.*

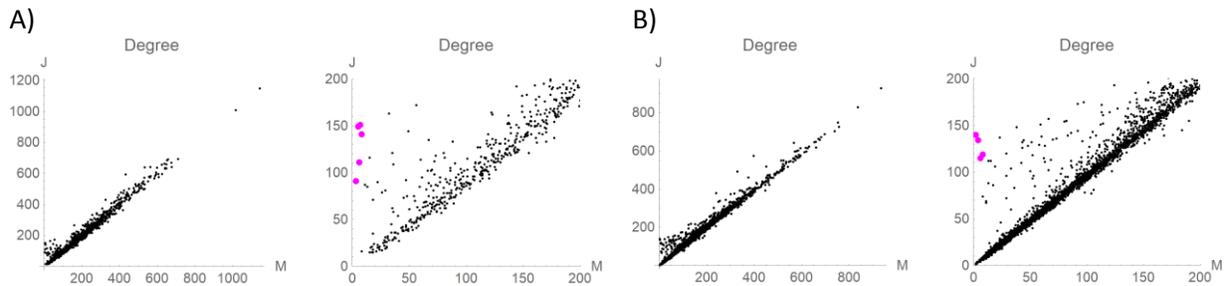

*Figure S 5: Number of interactions per genes (degree) detected by J and M. A) ExE library. B) NxN library. Zoom of the region degree<200 is shown in every case in the right plot. Pinks dots represent hubs trm112, tif35, noc4, rrp7 and tim17 from the ExE library and hubs zap1, vma7, msm1, and rpb4 from the NxN library.*

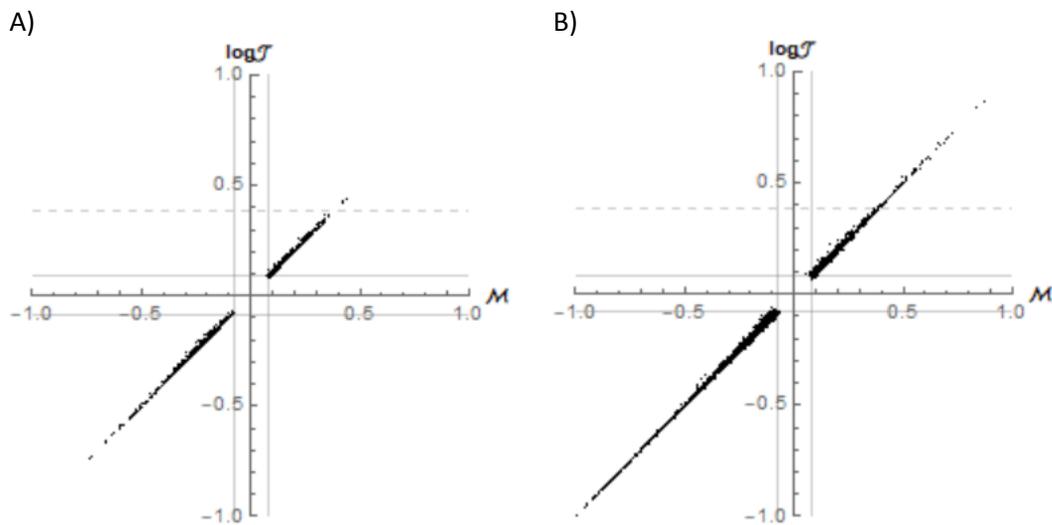

*Figure S 6: Comparison of the magnitude of interaction computed from the measure $\mathcal{M}$ (X axis), and the measure $\log \mathcal{J}$ (Y axis) of hubs similar in more than 95% of connections (i.e. $\overline{M} J + M \overline{J} < 0.05\, M J$, for $M J > 99$). Vertical and horizontal lines delimit the no-interaction region enclosed by $-0.08 < \mathcal{M} < 0.08$ and $-0.0886 < \log \mathcal{J} < 0.0886$. A) M vs log J plot of the shared hubs of ExE library. C) M vs log J plot of the shared hub of NxN library.*

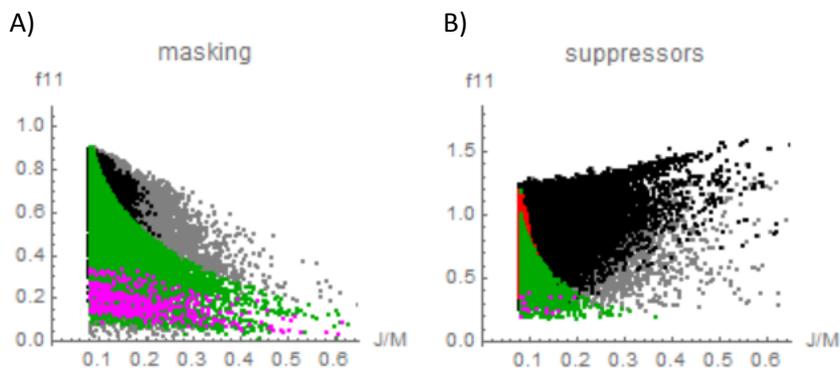

*Figure S 7: Double mutant fitness (f11) vs J/M interaction score plot comprising masking A) and suppressor B) positive interactions (NxN library). Black and dark-gray colors indicate MJ co-identified interactions (M, black; J, gray). $M \overline{J}$ (red dots) and $\overline{M} J$ (green dots). The newly identified hubs within the NxN library are highlighted in magenta.*

*Table S 1: Features of J-exclusive candidate Hubs within the ExE and NxN library (J>0 and M>0).*

*Data S 1:* **Segregation of J vs M interacting gene pairs (ExE library) across Biological Processes**

*Data S 2:* **Segregation of J vs M interacting gene pairs (ExE library) across Pathways**

*Data S 3:* **Segregation of J vs M interacting gene pairs (ExE library) across Molecular Complexes defined by Costanzo et al. 2016**

*Data S 4:* **Genes with expression profiles most similar to the query gene set rrp7, noc4 and trm112**

*Data S 5:* **Segregation of J vs M interacting gene pairs (NxN library) across Biological Processes**

*Data S 6:* **Segregation of J vs M interacting gene pairs (NxN library) across Pathways**

*Data S 7:* **Segregation of J vs M interacting gene pairs (NxN library) across Molecular Complexes defined by Costanzo et al. 2016**

*Data S 8: Hubs equally identified (shared) by J and M within the ExE and NxN libraries.*